\documentclass{article}

\usepackage{arxiv}

\usepackage[utf8]{inputenc} 
\usepackage[T1]{fontenc}    
\usepackage{hyperref}       
\usepackage{url}            
\usepackage{booktabs}       
\usepackage{amsfonts}       
\usepackage{nicefrac}       
\usepackage{microtype}      
\usepackage{graphicx}
\usepackage{natbib}
\usepackage{doi}

\usepackage{amsmath}
\usepackage{amssymb}
\usepackage{graphics}
\usepackage{plain}
\usepackage{charter}
\usepackage{xspace}
\usepackage[T1]{fontenc}
\usepackage[bitstream-charter]{mathdesign}

\usepackage{amsthm}
\newtheorem{theorem}{Theorem}[section]

\providecommand{\rk}[1]{\text{rank}\left(#1\right)}

\title{Mechanical normal form of first order state-space systems}

\author{ \href{https://orcid.org/0000-0002-2992-9038}{\includegraphics[scale=0.06]{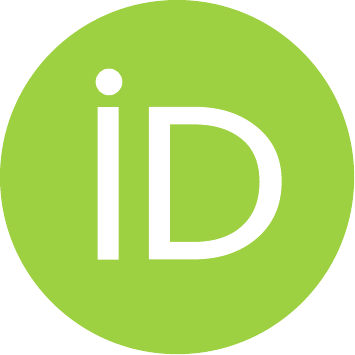}\hspace{1mm}Johannes~Mayet} \\
	Institute of Applied Mechanics\\
	Technical Unviversity Munich\\
	\texttt{johannesmayet@tum.de} \\
	\And
	\href{https://orcid.org/0000-0000-0000-0000}{\includegraphics[scale=0.06]{orcid.pdf}\hspace{1mm}Benjamin Kammermeier} \\
	Institute of Applied Mechanics\\
	Technical Unviversity Munich\\
	\texttt{benjamin.kammermeier@tum.de} \\
}

\renewcommand{\shorttitle}{Mechanical Normal Form}

\hypersetup{
pdftitle={Mechanical normal form of first order state space systems},
pdfauthor={Johannes Mayet, Benjamin Kammermeier},
pdfkeywords={Mechanical, Normal Form, Transformation},
}

\renewcommand{\vec}[1]{\boldsymbol{#1}}

\newcommand{\ode}{\textsf{ODE}\xspace}
\newcommand{\dof}{\textsf{DOF}\xspace}

\providecommand{\bmat}[1]{\begin{bmatrix} #1 \end{bmatrix}}

\begin{document}
\maketitle

\begin{abstract}
In this work a state transformation is presented that transforms a given state-space system to a normal form related to mechanical systems. The underlying state-space system must meet certain requirements such that a transformation exist. If the requirements are satisfied one obtains second order differential equations which allow the application of customized and specialized algorithms.
\end{abstract}


\section{Introduction}
It is very well known that state-space models can be realized in different standard forms
(e.g. Jordan (modal), observable and controllable canonical form). As described by~\cite{foellinger1978regelungstechnik} every canonical form offers some advantages for given tasks, e.g. the design of controllers is preferable carried out in controllable canonical form because changes of controller gains and their effect on the system behavior are made apparent. Furthermore, in the context of observability and controllabiltiy the state-space systems might be decomposed into four mutually exclusive parts with canonical structure~\citep{kalman1963canonical,kalman1963decomp}. In contrast the Jordan canonical form is best suited for theoretical aspects due to the decoupled state-space. \newline
In numerous applications a dynamical system has a given structure due to its origin, e.g. dynamics of electronic circuits. This underlying structure is often the strongest counter-argument to use  state-space systems 
as the physical structure is in general not retained, e.g. when calculating eigenvalues. Depending on the mathematical operation and field of application it may lead to dynamical behavior which is not in agreement with our physical models of nature. Furthermore, state-space systems equipped with additional physical knowledge may allow the use of customized and specialized approaches to system identification or structure preserving model order reduction~\citep{tulleken1994grey,wolf2009mor_passivity, wolf2014mor_passivity,freund2008mor_preserve,benner2016mor_2nd}. \newline
In this work the existence as well as the numerical solution method for a transformation, which yields a set of second order differential equations, is presented for a first order state-space system. Of course, such a transformation will only exist if the state-space system meets certain requirements.~\cite{salimbahrami2004folding} investigated the existence of such transformations for a single-input and single-output state-space system with restriction on the Markov parameters. Although the same objective is pursued, the approach is significantly different. Additionally, the proposed extension takes multiple-input and multiple-output systems into account. 
\section{Mechanical Normal Form}
The ordinary differential equation (\ode) of classical mechanical systems with $n/2$ degrees of freedom (\dof) represented by minimal coordinates $\vec{q}\in\mathbb{R}^{n/2\times 1}$ and velocities $\dot{\vec{q}}\in\mathbb{R}^{n/2\times 1}$ is given by
\begin{subequations}
\begin{align}
\vec{M}\vec{\ddot{q}} + \vec{h}(\vec{q},\vec{\dot{q}},t) &= \vec{W}\vec{u}\\
\vec{g}(\vec{q},\vec{\dot{q}},t)&=\vec{0}\,,
\end{align}
\label{equ:mech_sys_2nd_order}
\end{subequations}
where $\vec{W}\in\mathbb{R}^{n/2\times p}$ is the direction matrix of the $p$ external forces $\vec{u}\in\mathbb{R}^{p\times 1}$ (cf.~\cite{bremer1988mechsys}). All dynamic and static loads (e.g. external forces and torques) contained in $\vec{h}(\vec{q},\vec{\dot{q}},t)\in\mathbb{R}^{n/2\times 1}$ cause accelerations $\ddot{\vec{q}}\in\mathbb{R}^{n/2\times 1}$ weighted by the mass matrix $\vec{M}\in\mathbb{R}^{n/2\times n/2}$.
In this representation, the external forces $\vec{u}$ also include constraint forces needed to satisfy additional (non-holonomic) constraints $\vec{g}$. Although these kinematic constraint equations are formally taken into account, they will not be explicitly considered in this work. Since the inverse of the mass matrix $\vec{M}=\vec{M}^T>\vec{0}$ of such mechanical systems always exists, it is possible to rewrite the \ode as follows:
\begin{align}
\vec{\dot{x}} &= \frac{\mathrm{d}}{\mathrm{d}t}\bmat{\vec{q} \\ \vec{\dot{q}}} = 
\bmat{\vec{\dot{q}} \\ -\vec{M}^{-1}\vec{h}(\vec{x},t)} +  \bmat{\vec{0} \\ \vec{M}^{-1}\vec{W}}\vec{u}
\label{equ:mech_sys}
\end{align}
Linearization of Eq.~\eqref{equ:mech_sys} with $\vec{x}=\bmat{\vec{q} & \vec{\dot{q}}}$ then yields a first order differential equation with specific form:
\begin{align}
\vec{\dot{x}} & = 
\bmat{\vec{0} & \vec{I}\\ \vec{A}_1 & \vec{A}_2}\vec{x} +  \bmat{\vec{0} \\ \vec{B}_2}\vec{u}\,.
\label{equ:mech_sys_lin}
\end{align}
In many situations it is known from the beginning that the first order system, e.g. generated by a system identification algorithm, is a mechanical system and therefore can be represented by a set of second order differential equations given by Eq.~\eqref{equ:mech_sys_lin}. For this reason, the primary interest lies in clarifying under which conditions it is possible to obtain Eq.~\eqref{equ:mech_sys_lin} by applying an appropriate equivalence transformation to a given state-space system. A general linear time-invariant (LTI) state-space system is formally represented by
\begin{subequations}
\begin{align}
\vec{\dot{x}} &= \vec{A} \vec{x} + \vec{B} \vec{u}\\
\vec{y} &= \vec{C} \vec{x} + \vec{D} \vec{u}\,,
\end{align}
\label{equ:lti}
\end{subequations}
with state vector $\vec{x}\in\mathbb{R}^n$, output vector $\vec{y}\in\mathbb{R}^q$ and
input vector $\vec{u}\in\mathbb{R}^p$. In this case the state matrix $\vec{A}\in\mathbb{R}^{n \times n}$, input matrix $\vec{B}\in\mathbb{R}^{n \times p}$, output matrix $\vec{C}\in\mathbb{R}^{q \times n}$ and feed-through $\vec{D}\in\mathbb{R}^{q \times p}$ are assumed to be time-invariant. In order to be able to obtain Eq.~\eqref{equ:mech_sys_lin} by applying an equivalence transformation, one is evidently restricted to cases where the number of state variables $n$ is even, the number of inputs $p$ is equal or less than $n/2$, and the condition $\text{rank}(A) \geq n/2$. Without loss of generality it is assumed that the input matrix has full rank, that is $\text{rank}(B)=p$. Besides these obvious requirements it must be clarified in which situations the proposed transformation is well defined. Since the resulting matrices $\hat{\vec{A}}$ and $\hat{\vec{B}}$ must be of the form
\begin{align}
\hat{\vec{A}} & = \vec{T}\vec{A}\vec{T}^{-1} = \bmat{\vec{0} & \vec{I}\\ \hat{\vec{A}}_{21} & \hat{\vec{A}}_{22}} \quad \text{and}\quad \hat{\vec{B}} = \vec{T}\vec{B} 
= \bmat{\vec{0} \\ \hat{\vec{B}}_2}\,,
\label{equ:mech_sys_state_space}
\end{align}
it is required that the following statements are satisfied:
\begin{subequations}
\begin{align}
\bmat{\vec{I} & \vec{0}}\hat{\vec{A}} &= \bmat{\vec{0} & \vec{I}} & \rightarrow &&
\bmat{\vec{I} & \vec{0}}\vec{T}\vec{A} &= \bmat{\vec{0} & \vec{I}}\vec{T}
 \\
\bmat{\vec{I} & \vec{0}}\hat{\vec{B}} &= \vec{0} & \rightarrow &&
\bmat{\vec{I} & \vec{0}}\vec{T}\vec{B} &= \vec{0}
\end{align}
\label{equ:2nd_cond}
\end{subequations}
If the transformation matrix $\vec{T}^T=\bmat{\vec{X}  &  \vec{Y}}$ is split into two non-square matrices, then Eq.~\eqref{equ:2nd_cond} is equivalent to $\vec{Y}=\vec{A}^T\vec{X}$ and $\vec{B}^T\vec{X}=\vec{0}$ for any $\vec{X},\vec{Y}\in\mathbb{R}^{n \times n/2 }$ with $\text{rank}\left(\bmat{\vec{X}  &  \vec{Y}}\right) = n$. 
By introducing the orthonormal complement $\vec{X}_\bot$ ($\vec{X}_\bot^T\vec{X}_\bot=\vec{I}$ and $\vec{X}^T\vec{X}_\bot=\vec{0}$) it is shown that the equivalence transformation and associated inverse defined by
\begin{align}
\vec{T}^T&=\bmat{\vec{X}, &  \vec{A}^T\vec{X}} \\
\vec{T}^{-1} &= \bmat{\vec{X}-\vec{X}_\bot\left(\vec{X}^T\vec{A}\vec{X}_\bot\right)^{-1}\left(\vec{X}^T\vec{A}\vec{X}\right), & \vec{X}_\bot\left(\vec{X}^T\vec{A}\vec{X}_\bot\right)^{-1}} \,,
\end{align}
with $\vec{T}\vec{T}^{-1}=\vec{I}$ and $\det(\vec{T})=\det\left(\vec{X}^T\vec{A}\vec{X}_\bot\right)\neq0$ (cf. App.~\ref{app:detT}) transforms the original state-space system as required. From the definitions of the orthonormal matrices it is directly apparent that the question whether such a transformation exists can only be answered if one considers the interaction of the system matrix $\vec{A}$ and a subspace of the input matrix nullspace. Since $\vec{X}$ spans a subspace of the nullspace of $\vec{B}$, it might be the case that one specific choice of subspace yields a non-singular transformation matrix while others do not. Since a pure trial and error is certainly unsatisfactory, conditions must be specified on the basis of the matrices $\vec{A}$ and $\vec{B}$ in which case a non-singular transformation can be found. If the input matrix is factorized by a singular value decomposition 
\begin{align}
\vec{B} = \bmat{\vec{U}_{B,1} & \vec{U}_{B,2}}\bmat{\vec{\Sigma}_{B} \\ \vec{0}}\vec{V}_B^T\,,
\label{equ:svd_B}
\end{align}
with $\vec{U}_{B,1}\in \mathbb{R}^{n \times p}$, $\vec{U}_{B,2}\in \mathbb{R}^{n \times n-p}$, $\vec{\Sigma}_B\in \mathbb{R}^{n-p \times p}$ and $\vec{V}_B\in \mathbb{R}^{p \times p}$, it is possible to determine non-unique matrices $\vec{X}=\vec{U}_{B,2}\vec{E}$ and $\vec{X}_\bot=\bmat{\vec{U}_{B,1} & \vec{U}_{B,2}\vec{E}_\bot}$, where the columns of matrix $\vec{E}\in\mathbb{R}^{n-p \times n/2}$ are $n/2$ orthonormal basis vectors. The matrix $\vec{E}_\bot\in\mathbb{R}^{n-p \times n/2-p}$ is the associated complement of matrix $\vec{E}$. The main reason for introducing the matrix $\vec{E}$ is that $n/2-p$ left-singular vectors of the left null space of $\vec{B}$ can be swapped between the matrices $\vec{X}$ and $\vec{X}_\bot$ without violating one of the definitions. As already mentioned it is important to maintain this freedom since the remaining requirement
$\rk{\vec{X}^T\vec{A}\vec{X}_\bot}=n/2$ is not necessarily satisfied for every possible subspace, see App.~\ref{subsec:example}.
\subsection{Candidate for matrix \texorpdfstring{$\vec{E}$}{TEXT}}
With the definition of the non-unique matrices $\vec{X}=\vec{U}_{B,2}\vec{E}$ and $\vec{X}_\bot=\bmat{\vec{U}_{B,1} & \vec{U}_{B,2}\vec{E}_\bot}$ it is required that the nullspace of $\vec{X}^T\vec{A}\vec{X}_\bot$ is trivial:
\begin{align}
\vec{X}^T\vec{A}\vec{X}_\bot \vec{h} = \vec{E}^T
\bmat{
\vec{K} & \vec{U}_{B,2}^T\vec{A}\vec{U}_{B,2}\vec{E}_\bot
}\vec{h} \neq \vec{0}\quad \forall\vec{h}\in\mathbb{R}^{n/2\times 1}
\text{ with } \vec{K} = \vec{U}_{B,2}^T\vec{A}\vec{U}_{B,1}
\label{equ:trivial_nullspace}
\end{align}
If vector $\vec{h}$ is divided into $\vec{h}_1 \in\mathbb{R}^{p\times 1}$ and $\vec{h}_2 \in\mathbb{R}^{n/2-p\times 1}$ it is apparent that matrix $\vec{K} = \vec{U}_{B,2}^T\vec{A}\vec{U}_{B,1}=\bmat{\vec{U}_{K,1}& \vec{U}_{K,2}}\bmat{\vec{\Sigma}_K^T & \vec{0}}^T \vec{V}_K^T$ must have rank $p$. Furthermore, matrix $\vec{E}=\bmat{\vec{U}_{K,1} & \vec{U}_{K,2}\vec{Z}}$ and $\vec{E}_\bot=\vec{U}_{K,2}\vec{Z}_\bot$ with $\vec{Z}^T\vec{Z}_\bot=\vec{0}$ for matrices $\vec{Z}\in\mathbb{R}^{n-2p\times n/2-p}$ and $\vec{Z}_\bot\in\mathbb{R}^{n-2p\times n/2-p}$ are suitable choices, which yields
\begin{align}
\vec{X}^T\vec{A}\vec{X}_\bot = 
\bmat{
\vec{\Sigma}_K\vec{V}_K^T & \vec{U}_{K,1}^T\vec{U}_{B,2}^T\vec{A}\vec{U}_{B,2}\vec{U}_{K,2}\vec{Z}_\bot \\
\vec{0}
& \vec{Z}^T\vec{U}_{K,2}^T\vec{U}_{B,2}^T\vec{A}\vec{U}_{B,2}\vec{U}_{K,2}\vec{Z}_\bot 
}\,.
\end{align}
Therefore, the matrices $\vec{Z}$ and $\vec{Z}_\bot$ must be specified such that the resulting square matrix $\vec{Z}^T\vec{G}\vec{Z}_\bot$ with $\vec{G}=\vec{U}_{K,2}^T\vec{U}_{B,2}^T\vec{A}\vec{U}_{B,2}\vec{U}_{K,2}\in\mathbb{R}^{n-2p\times n-2p}$ is invertible. As a direct consequence it has to be required that matrix $\vec{G}$ satisfies $\rk{\vec{G}}\geq n/2-p$. Let $\vec{G}=\vec{V}\vec{J}\vec{V}^{-1}$ be represented by the real Jordan normal form and define $\bar{\vec{Z}}=\vec{V}^T\vec{Z}
\in\mathbb{R}^{n-2p\times n/2-p}$, $\bar{\vec{Z}}_\bot
=\vec{V}^{-1}\vec{Z}_\bot\in\mathbb{R}^{n-2p\times n/2-p}$ in order to obtain $\vec{Z}^T\vec{G}\vec{Z}_\bot = \bar{\vec{Z}}^T\vec{J}\bar{\vec{Z}}_\bot$. Since $\vec{J}=\text{blkdiag}\left(\vec{J}_1,\vec{J}_2,\cdots\right)$ is a block diagonal matrix where every Jordan block is of the form
\begin{align}
\vec{J}_i=
\bmat{
\vec{C}_i & \vec{I}   &  		&  		    \\
          & \vec{C}_i & \ddots 	& 		    \\
          &           & \ddots 	& \vec{I}   \\
          &           &        	& \vec{C}_i \\
}\,,
\label{equ:jordan_block_cpl}
\end{align}
with $\vec{C}_i = \bmat{a_i & -b_i \\ b_i & \phantom{-} a_i}$ for complex conjugate eigenvalue pairs $\lambda = a_i \pm ib_i$ or $\vec{C}_i = \lambda_i$ for real eigenvalues\footnote{The dimension of the identity matrix is equal to the dimension of $\vec{C}_i$.}, the unknown matrices $\vec{Z}$ and $\vec{Z}_\bot$ can be determined block-wise. As it is shown in App.~\ref{app:matP} it is straight forward to find matrices $\bar{\vec{Z}}$ and $\bar{\vec{Z}}_\bot$ such that $\vec{Z}^T\vec{Z}_\bot=\vec{0}$ and $\rk{\bar{\vec{Z}}^T\vec{J}\bar{\vec{Z}}_\bot} = n/2-p$ if and only if $\rk{\vec{G}}\geq n/2-p$. As a consequence the matrix $\vec{E}$ defined by $\vec{E}=\bmat{\vec{U}_{K,1} & \vec{U}_{K,2}\left(\vec{V}^T\right)^{-1}\bar{\vec{Z}}}$ is a valid candidate.
\newline
Therefore, a non-singular equivalence transformation $\vec{T}^T=\bmat{\vec{X}, &  \vec{A}^T\vec{X}}^T$  with $\vec{X}=\vec{U}_{B,2}\bmat{\vec{U}_{K,1} & \vec{U}_{K,2}\left(\vec{V}^T\right)^{-1}\bar{\vec{Z}}}$ for an arbitrary mechanical system is completely defined. Yet with the intention of a uniform representation of a mechanical system, a further transformation is to be defined.
\subsection{Subsequent Unifying Transformation}
Since the matrix $\vec{B}_2$ of Eq.~\eqref{equ:mech_sys_state_space} has rank $p$, one may additionally apply a state transformation $\tilde{\vec{x}} = \vec{W} \vec{\hat{x}}$ where $\vec{W} = \text{blkdiag}(\vec{W}_1,\vec{W}_1)$ such that the state-space system $\vec{\dot{\tilde{x}}} = \vec{\tilde{A}}\vec{\tilde{x}} + \vec{\tilde{B}}\vec{u}$  and $\vec{y} = \vec{\tilde{C}}\vec{\tilde{x}} + \vec{\tilde{D}}\vec{u}$ with matrices of the following form is obtained:
\begin{align}
\vec{\tilde{D}} &= \vec{D} \qquad
\vec{\tilde{C}} = \vec{C}\left(\vec{W}\vec{T}\right)^{-1} \\
\vec{\tilde{B}} &=  \left(\vec{W}\vec{T}\right)\vec{B} 
= \bmat{\vec{0}_{n/2-p \times p} \\ \vec{I}_{p \times p}} 
\qquad
\vec{\tilde{A}} = \left(\vec{W}\vec{T}\right)\vec{A}\left(\vec{W}\vec{T}\right)^{-1} = \bmat{\vec{0} & \vec{I}\\ \vec{\tilde{A}}_{21} & \vec{\tilde{A}}_{22}}
\end{align}
This subsequent equivalence transformation is in general not unique. One way to show this is to apply another subsequent transformation $\vec{Q} = \text{blkdiag}(\vec{Q}_1,\vec{Q}_1)$ that leaves the input matrix and the upper $n/2$ rows of the state matrix invariant. Mathematically this yields 
$\vec{Q}_1\bmat{\vec{0} & \vec{I}}^T = \bmat{\vec{0} & \vec{I}}^T$, which is satisfied for any invertible matrix $\vec{Q}_1 = \bmat{\vec{Q}_{11} & \vec{0}
\\ \vec{Q}_{12} & \vec{I}} $ with $\vec{Q}_{11}\in\mathbb{R}^{n/2-p \times n/2 - p}$ and $\vec{Q}_{12}\in\mathbb{R}^{p \times n/2 - p}$ for $p<n/2$. Invertibility of $\vec{Q}_1$ is guaranteed by $\vec{Q}_{11}$ being invertible. In case $p=n/2$, the proposed equivalence transformation is indeed unique, and consequently $\vec{Q}_1=\vec{I}$. The subsequent transformation is of course not necessary, but it serves as a unification for the mechanical normal form. The following theorem summarizes the findings of this contribution:
\begin{theorem}[Mechanical Normal Form]
A linear time-invariant state-space system given in Eq.~\eqref{equ:lti} with an even number of states $n$, $p\leq n/2$ numbers of inputs, and $\text{rank}(B)=p$ can be represented by an inhomogeneous second order differential equation  with $n/2$ number of states (cf. Eq.~\eqref{equ:mech_sys_2nd_order}) by applying an equivalence state transformation $\vec{T}^T=\bmat{\vec{X}, &  \vec{A}^T\vec{X}}$, where the orthonormal matrix $\vec{X}\in\mathbb{R}^{n\times n/2}$ with $\vec{X}^T\vec{X}=\vec{I}$ satisfies $\vec{B}^T\vec{X}=\vec{0}$ and $\text{rank}\left(\bmat{\vec{X}  &  \vec{A}^T\vec{X}}\right) = n$. A non-singular transformation exists iff
$\rk{\vec{U}_{B,2}^T\vec{A}\vec{U}_{B,1}}=p$ and $\rk{\left(\vec{U}_{B,2}\vec{U}_{K,2}\right)^T\vec{A}\left(\vec{U}_{B,2}\vec{U}_{K,2}\right)}\geq n/2-p$ are satisfied, where $\vec{U}_{B,2}$ resp. $\vec{U}_{K,2}$ is an orthonormal basis of the nullspace of matrix $\vec{B}$ resp. $\vec{K}=\vec{U}_{B,2}^T\vec{A}\vec{U}_{B,1}$, and the columns of matrix $\vec{U}_{B,1}$ are the first $p$ left-singular vectors of matrix $\vec{B}$. A given linear time-invariant system $\left(\vec{A},\vec{B},\vec{C},\vec{D}\right)$ satisfying these conditions can be represented in mechanical normal form $\left(\vec{\tilde{A}},\vec{\tilde{B}},\vec{\tilde{C}},\vec{\tilde{D}}\right)$ with $\bmat{\vec{I} & \vec{0}}\vec{\tilde{A}} = \bmat{\vec{0} & \vec{I}}$ and $\vec{\tilde{B}} = \bmat{\vec{0} & \vec{I}}^T$.
\label{thm:mechform}
\end{theorem}

\section{Numerical Implementation}\label{sec:num_impl}
Although the above sections describe one way to determine the matrix $\vec{E}$, it will in most applications not be the preferred approach due to numerical accuracy and efficiency. As it is possible to rewrite the mechanical normal form requirements by $\bmat{\vec{I} & \vec{0}}\vec{T}\vec{A} = \bmat{\vec{0} & \vec{I}}\vec{T}\vec{I}$ and $\vec{T}\vec{B}\vec{I} = \bmat{\vec{0} & \vec{I}}^T$ of Thm.~\ref{thm:mechform}, one obtains a single matrix equation $\vec{L}\,\text{vec}(\vec{T}) = \vec{R}$ using vectorization and the kronecker product given by
\begin{align}
\vec{R} = \bmat{\vec{0} & \text{vec}\left(\bmat{\vec{0} & \vec{I}}^T\right)^T
}^T
\text{ and }
\vec{L} =
\bmat{
\vec{A}^T\otimes\bmat{\vec{I} & \vec{0}}-\vec{I}\otimes
\bmat{\vec{0} & \vec{I}} \\
\vec{B}^T\otimes \vec{I}
}\label{equ:num_impl_L}
\,.
\end{align}

The solution can for example be obtained by singular-value decomposition, $\vec{T}_0$ is then obtained by devectorization of $\text{vec}(\vec{T}_0)$. However, this initial solution might be a singular matrix. This can be circumvented by determining the nullspace of $\vec{L}$ from which one obtains matrices $\vec{T}_1,\vec{T}_2,\cdots,\vec{T}_s$ defined by $\bmat{\text{vec}(\vec{T}_1) & \text{vec}(\vec{T}_2) & \cdots & \text{vec}(\vec{T}_s)} = \text{null}\left(\vec{L}\right)\in\mathbb{R}^{n^2\times s}$. These matrices are then used to construct the transformation matrix $\vec{T} = \vec{T}_0 + \sum_{k=1}^s \alpha_k \vec{T}_k$ as a sum of matrices with unknown coefficients $\alpha_1,\alpha_2,\cdots,\alpha_s\in\mathbb{R}$. In this work it is proposed to generate $n$ independent column vectors $\vec{t}_k$ of $\vec{T} = \bmat{\vec{t}_1  &  \vec{t}_2  & \cdots &  \vec{t}_n }$ by minimizing the cost function $J(\alpha_1,\alpha_2,\cdots,\alpha_s)=\sum_{k=1}^n\sum_{j=1,j\neq k}^n |\vec{t}_k^T\vec{t}_j| / \left(|\vec{t}_k| |\vec{t}_j|\right)$. In general it will not be possible to guarantee a solution that yields an invertible transformation matrix. But in case that all requirements of Thm.~\ref{thm:mechform} are satisfied, it is known that a solution must exist. Depending on the optimization algorithm one might be concerned about convergence and local minima. However, since it is only required that the matrix $\vec{T}$ has full rank these considerations have minor importance.   
\subsection{Extension for descriptor state-space systems}
In many applications the state-space system is given in descriptor form $\vec{F}\vec{\dot{x}} = \vec{A}\vec{x} + \vec{B}\vec{u}$. With the assumption of this work that matrix $\vec{F}$ is invertible, the underlying equations of Thm.~\ref{thm:mechform} are given by
$\bmat{\vec{I} & \vec{0}}\vec{T}\vec{A} = \bmat{\vec{0} & \vec{I}}\vec{T}\vec{F}$ and $\vec{T}\vec{B}\vec{I} = \bmat{\vec{0} & \vec{I}}^T$. Therefore, modification of matrix $\vec{L}$ from Eq.~\eqref{equ:num_impl_L} given by\begin{align}
\vec{L} =
\bmat{
\vec{A}^T\otimes\bmat{\vec{I} & \vec{0}}-\vec{F}\otimes
\bmat{\vec{0} & \vec{I}} \\
\vec{B}^T\otimes \vec{I}}\,,
\end{align}
allows to consider descriptor systems without inverting the "mass matrix" $\vec{F}$.
\section{Conclusion}
In this work an equivalence transformation is derived that allows one to represent a linear time-invariant state-space system as a set of second order differential systems related to mechanical systems. Necessary conditions for the underlying state-space system are given and a numerical implementation is proposed.\newline
Future research work will focus on descriptor state-space systems with singular mass matrix and the integration of model order reduction procedures.
\appendix
\section{Determinant of Transformation Matrix \texorpdfstring{$\vec{T}$}{TEXT}}
\label{app:detT}
Since the determinant is a multiplicative map and it holds that $\det{\left(\bmat{\vec{X}_\bot &\vec{X}}\right)}=1$, the determinant of the transformation matrix $\vec{T}\in\mathbb{R}^{n\times n}$ is given by 
\begin{subequations}
\begin{align}
\det{(T)}&=\det{\left(\bmat{\vec{X}^T \\ \vec{X}^T \vec{A}}\right)} = \det{\left(\bmat{\vec{X}^T \\ \vec{X}^T \vec{A}}\right)}\det{\left(\bmat{\vec{X}_\bot &\vec{X}}\right)}\\
&=\det{\left(\bmat{\vec{0} & \vec{I} \\ \vec{X}^T\vec{A}\vec{X}_\bot & \vec{X}^T \vec{A}\vec{X}}\right)} = \det{\left(\vec{X}^T\vec{A}\vec{X}_\bot\right)}\,.
\end{align}
\end{subequations}
Note that the dimension of matrix $\vec{X}^T\vec{A}\vec{X}_\bot\in\mathbb{R}^{n/2 \times n/2}$ is half of the dimension of transformation $\vec{T}$.
\section{Example for a Singular Transformation Matrix \texorpdfstring{$\vec{T}$}{TEXT}}
In this example a system which is already represented by matrices $\vec{A}$ and $\vec{B}$ having the desired form is addressed:
\label{subsec:example}
\begin{align}
\vec{A} = \bmat{0 & 0 & 1 & 0 \\ 
		        0 & 0 & 0 & 1 \\
		        -\omega_1^2 & 0 & 0 & 0 \\
		        0 & -\omega_2^2 & 0 & 0 
             }
             \quad
\vec{B}=\bmat{0\\0\\0\\1} \quad \rightarrow \quad
\vec{U}_{B,2} = \bmat{1 & 0 & 0 \\0 & 1 & 0 \\0 & 0 & 1\\ 0 & 0 & 0}
\end{align}
Assuming it is not known that the given system has already the desired form, one may choose the first and last left-singular vector of $\vec{B}$ and obtains
\begin{align}
\vec{X} = \bmat{1 & 0 & 0 & 0 \\ 0 & 0 & 1 & 0}^T \quad \rightarrow \quad 
\vec{T}&=\bmat{\vec{X}, &  \vec{A}^T\vec{X}}
=
\bmat{1 & 0 & 0 & -\omega_1^2\\
      0 & 0 & 0 & 0 \\
      0 & 1 & 1 & 0 \\
      0 & 0 & 0 & 0}\,,
\end{align}
which is obviously a singular transformation matrix. On the other hand it is easily verified that $\vec{X} = \bmat{\vec{I} & \vec{0}}^T$ (first and second left-singular vector) yields $\vec{T}^T=\bmat{\vec{X}, &  \vec{A}^T\vec{X}}^T
= \vec{I}$ as expected.
\section{Orthonormal Matrices \texorpdfstring{$\bar{\vec{Z}}$ and $\bar{\vec{Z}}_\bot$}{TEXT}}\label{app:matP}
Since $\vec{J}$ is a block-diagonal matrix it is possible to determine the vectors of $\bar{\vec{Z}}$ and $\bar{\vec{Z}}_\bot$ for every Jordan block separately. The following four cases show how the matrices can be constructed:
\subsection{Case 1: Conjugate Complex Eigenvalue}
Since a Jordan block $\vec{J}_i$ of a complex conjugate eigenvalue pair $\lambda_i = a_i \pm b_i$ with multiplicity is given by
Eq.~\eqref{equ:jordan_block_cpl} with $\vec{C}_i = \bmat{a_i & -b_i \\ b_i & \phantom{-} a_i}$, it can be verified that $\bar{\vec{Z}} = \frac{1}{\sqrt{a_i^2 + b_i^2}}\text{blkdiag}\left(\vec{c}_i,\vec{c}_i,\cdots,\vec{c}_i\right)$ and 
$\bar{\vec{Z}}_\bot = \frac{1}{\sqrt{a_i^2 + b_i^2}}\text{blkdiag}\left(\vec{c}_{i,\bot},\vec{c}_{i,\bot},\cdots,\vec{c}_{i,\bot}\right)$ 
with $\vec{c}_{i}=\bmat{a_i & b_i}^T$ and $\vec{c}_{i,\bot}=\bmat{b_i & - a_i}^T$ are suitable matrices. As a result one directly obtains $\bar{\vec{Z}}^T\bar{\vec{Z}}_\bot = \vec{0}$, $ \bar{\vec{Z}}^T\bar{\vec{Z}}= \left(\bar{\vec{Z}}_\bot\right)^T\bar{\vec{Z}}_\bot= \vec{I}$ and $\bar{\vec{Z}}^T\vec{J}_i\bar{\vec{Z}}_\bot =b_i\vec{I}$.
\subsection{Case 2: Real Eigenvalues}
In case that the Jordan matrix is given by $\vec{J}=\bmat{\lambda_1 & 0 \\ 0 & \lambda_2}$, where $\lambda_1\neq0$ and $\lambda_2$ are unequal real eigenvalues, one may take $\bar{\vec{Z}} = \frac{\sqrt{2}}{2}\bmat{1 & 1}^T$ and $\bar{\vec{Z}}_\bot = \frac{\sqrt{2}}{2}\bmat{1 & -1}^T$ with $\bar{\vec{Z}}^T\vec{J}_i\bar{\vec{Z}}_\bot =\lambda_1-\lambda_2\neq0$. Note that $\lambda_2$ is allowed to be zero.
\subsection{Case 3: Real Eigenvalue with Even Multiplicity}
In case that the Jordan block matrix is given by $\vec{J}_i=\bmat{\lambda_i & 1 \\ 0 & \lambda_i}$ with $\lambda_1\neq0$, one may take $\bar{\vec{Z}} = \bmat{1 & 0}^T$ and $\bar{\vec{Z}}_\bot = \bmat{0 & 1}^T$ with $\bar{\vec{Z}}^T\vec{J}_i\bar{\vec{Z}}_\bot = 1$.
\subsection{Case 4: Real Eigenvalues with Odd Multiplicity}
In case that the Jordan matrix is given by 
\begin{align}
\vec{J}=\bmat{\lambda_1 & 1 & 0 & 0 \\ 0 & \lambda_1 & 1 & 0 \\ 0 & 0 & \lambda_1 & 0 \\ 0 & 0 & 0 & \lambda_2}\,,
\end{align}
where $\lambda_1\neq0$ with odd multiplicity and $\lambda_2\neq\lambda_1$, the orthonormal matrices $\bar{\vec{Z}} = \frac{1}{\sqrt{2}} \bmat{\sqrt{2} & 0 & 0 & 0 \\ 0 & 0 & 1 & 1}^T$ and $\bar{\vec{Z}}_\bot = \frac{1}{\sqrt{2}} \bmat{0 & \sqrt{2} & 0 & \phantom{-}0 \\ 0 & 0 & 1 & -1}^T$ yield $\bar{\vec{Z}}^T\vec{J}\bar{\vec{Z}}_\bot = \bmat{1 & 0 \\ 0 & \lambda_1-\lambda_2}$.
\subsection{Multiple Eigenvalues Equal Zero}
As shown by the four different cases it is always possible to find orthonormal matrices $\bar{\vec{Z}}$ and $\bar{\vec{Z}}_\bot$ for given Jordan blocks with the assumption that at least one real eigenvalue is not equal zero. Due to the fact that matrix $\vec{G}\in\mathbb{R}^{n-2p\times n-2p}$ may have multiple zero eigenvalues (requirement $\rk{\vec{G}}\geq n/2 - p$), not all possible scenarios are already covered. However, without loss of generality the block-diagonal Jordan matrix can be arranged such that an even number $f\leq n/2 - p$ of zero eigenvalues are gathered in the last block matrix, e.g. $\vec{J}=\text{blkdiag}\left(\bar{\vec{J}}, \vec{0}_{f\times f}\right)$. Note that the dimension of $\bar{\vec{J}}$ is always even, therefore, the four cases completely cover the Jordan matrix $\bar{\vec{J}}\in\mathbb{R}^{n-f-2p \times n-f-2p}$ even if an odd number of zero-eigenvalues exist. If we denote the columns of $\bar{\vec{Z}}$ and $\bar{\vec{Z}}_\bot$ with
$\bar{\vec{z}}_1,\bar{\vec{z}}_2,\cdots,\bar{\vec{z}}_{(n-f)/2-p}$ and $\bar{\vec{z}}_{1,\bot},\bar{\vec{z}}_{2,\bot},\cdots,\bar{\vec{z}}_{(n-f)/2-p,\bot}$ associated to the decoupled Jordan blocks of the Jordan matrix $\bar{\vec{J}}$, then it is possible to construct the remaining columns by simple extensions. In case of $f=2$ the columns read as follows:

\begin{align}
\bar{\vec{Z}} =
\bmat{
\bar{\vec{z}}_1 &     &  & \vec{0} & \vline & \bar{\vec{z}}_1 \\
				&  \ddots & & & \vline &  \vec{0}\\
				& &\ddots & & \vline &  \vdots \\
\vec{0}  & & & \bar{\vec{z}}_{(n-f)/2-p} & \vline & \vec{0} \\			
\hline
1 & 0 & \cdots & 0 & \vline & -\bar{\vec{z}}_1^T\bar{\vec{z}}_1 \\
0 & 0 & \cdots & 0 & \vline & 0 }
&&
\bar{\vec{Z}}_\bot =
\bmat{
\bar{\vec{z}}_{1,\bot} &     &  & \vec{0} & \vline & \bar{\vec{z}}_{1,\bot} \\
				&  \ddots & & & \vline &  \vec{0}\\
				& &\ddots & & \vline &  \vdots \\
\vec{0}  & & & \bar{\vec{z}}_{(n-f)/2-p,\bot} & \vline & \vec{0} \\			
\hline
0 & 0 & \cdots & 0 & \vline & 0 \\
1 & 0 & \cdots & 0 & \vline & -(\bar{\vec{z}}_{1,\bot})^T\bar{\vec{z}}_{1,\bot} 
}
\end{align}
In a last step the columns of the matrices can be scaled to length 1 in order to obtain orthonormal matrices. In summary, the matrix $\bar{\vec{Z}}^T\vec{G}\bar{\vec{Z}}_\bot$ will be a diagonal matrix with entries unequal zero and therefore the rank of this matrix is equal $n/2-p$ as required. Note that the presented procedure is certainly not unique, which allows further considerations to be addressed, e.g. numerical accuracy and computation time.
\bibliographystyle{unsrtnat}
\bibliography{normal_form}  

\end{document}